\documentclass[twocolumn,showpacs,amsmath,amssymb,prb,graphics,graphicx]{revtex4-1}
     \usepackage{graphicx}
\usepackage{bm}
\usepackage{dcolumn}
\begin{document}

\title{Time dependent London approach, dissipation due to out-of-core normal excitations by moving vortices}

\author{ V. G. Kogan }
\affiliation{  Ames Laboratory-DOE and Department of
Physics, Iowa State University, Ames, Iowa 50011 }
      \date{\today}
\begin{abstract}
The dissipative currents due to normal excitations are included in the
London description. The resulting time dependent London equations are solved
for a moving vortex and a moving vortex lattice.
It is shown that the field distribution of a moving vortex looses it cylindrical symmetry, it experiences  contraction which is stronger in the direction of the motion, than in the direction normal to the velocity $\bm v$. 
The London contribution of normal currents to dissipation is small relative to the Bardeen-Stephen core dissipation at small velocities, but approaches the latter at high velocities, where this contribution is no longer proportional to $v^2$. To minimize the London contribution to dissipation, the vortex lattice orients as to have one of the unit cell vectors along the velocity, the effect seen in experiments and predicted within the time-dependent Ginzburg-Landau theory. 
  \end{abstract}

\noindent 

\maketitle

\section{Introduction}

The London (L) equations have been proved to be a useful tool in describing magnetic properties of  superconductors. Originally, they were obtained by brothers London using a heuristic argument to describe the Meissner effect. \cite{London} Later they were derived from the microscopic theory for all temperatures or, for temperatures near the critical, from the Ginzburg-Landau (GL) current expression, 
\begin{equation}
{\bm J}=  -\frac{2e^2 f^2}{mc}\, \left( {\bm A}+\frac{\phi_0}{2\pi}{\bm \nabla}\chi\right)  \,,
\label{currentGL}
\end{equation}
 by setting constant the order parameter modulus $f$. This is the major model shortcoming: the cores of vortices cannot be described by the L-theory ($\bm A$ is the vector potential,   $\phi_0$ is the flux quantum, and $\chi$ is the phase). Applying curl to Eq.\,(\ref{currentGL}) one obtains L-equations 
\begin{equation}
{\rm curl\,}{\rm curl\,}{\bm H}+\frac{1}{\lambda^2}\,{\bm
H} =\frac{\phi_0}{\lambda^2}\hat{\bm z}\delta({\bm r}-{\bm r_\nu})\,,\label{L}
\end{equation}
where ${\bm r_\nu} $ is the position of the phase singularity and $ \hat {\bm z}$ is the straight vortex direction. For more than one vortex, the right-hand side here contains a sum of delta-functions. Notwithstanding this derivation from GL theory,   the L-equations hold for all temperatures since they, in fact, express the Meissner effect (under the caveat mentioned).   Of course, the temperature and scattering dependences of the penetration depth $\lambda$, the only material parameter of the L-theory, should be taken from a microscopic consideration or from experiment.

A linear Eq.\,(\ref{L}) (or its anisotropic version) is instrumental in studying static  intervortex interactions and  vortex lattices (VL)  for which  vortex cores are  irrelevant provided the intervortex spacing $a\gg\xi$, the coherence length or the core size.

It this article, the time dependent L-equations are discussed. This approach was employed in literature for quite some time,  see e.g. Ref.\,\onlinecite{Clem-Coffey}. Here, 
it is applied   to moving vortices and vortex lattices (VLs). Moving VLs have been  studied  in a number of experiments.\cite{Yaron,March,Pardo,Forgan}  
It was shown theoretically that in the presence of pinning the vortex system does not have a long range order at low velocities, but  orders to a VL at high velocities with one of the lattice vectors along the velocity.\cite{Kosh-Vin} This VL orientation was also proven to be preferable in the absence of pinning near  $T_c$   with the help of time-dependent GL equations (TDGL).\cite{Barukh}
As is shown below, the time-dependent London approach (TDL) provides a simpler way to address, e.g., the question of structure of moving VLs, or in general, the $t$-dependent problems in which   vortex cores do not play a role.


      In   time dependent situations,   the current   consists, in
general, of  normal and superconducting parts:
\begin{equation}
{\bm J}= \sigma {\bm E} -\frac{2e^2 f^2}{mc}\, \left( {\bm
A}+\frac{\phi_0}{2\pi}{\bm
\nabla}\chi\right)  \,,\label{current}
\end{equation}
where ${\bm E}$ is the electric field.    The conductivity  $\sigma$    for the quasiparticles flow  is in
general frequency  dependent.\cite{Tinkham} If however the frequencies   are bound by
inequality $\omega\tau_n\ll 1$ with $\tau_n$ being the scattering time for the
normal excitations, one can consider $\sigma$ as a real
$\omega$-independent quantity. Usually, the term with the normal conductivity is small because the density of normal excitations $\rho_n$ is negligible away of $T_c$ for s-wave materials. However, close to $T_c$ or in gapless materials, $\rho_n $ is practically close to that of the normal phase.

Within the London model $f^2=$ const and Eq.\,(\ref{current}) becomes:
\begin{equation}
\frac{4\pi}{c}{\bm J}= \frac{4\pi\sigma}{c} {\bm E} -\frac{1}{\lambda^2}\,
\left( {\bm A}+\frac{\phi_0}{2\pi}{\bm \nabla}\chi\right)  \,.
\label{current1}
\end{equation}
  Operating by curl one obtains:
\begin{equation}
{\rm curl\,}{\rm curl\,}{\bm H}+\frac{1}{\lambda^2}\,{\bm
H}+\frac{4\pi\sigma}{c^2}\,\frac{\partial {\bm H}}{\partial
t}=\frac{\phi_0}{\lambda^2}{\bm z}\sum_{\bm r_\nu} \delta({\bm r}-{\bm r_\nu})\,,\label{TDL}
\end{equation}
where ${\bm r_\nu}(t) $ are positions of the phase singularities which might change in time. 

\section{Moving vortex  }

Equation (\ref{TDL}) can be considered as a general form of the time
dependent London equation. 
For a straight vortex along ${\bm z}$ moving with
a constant velocity ${\bm v}$ in the $xy$ plane, this equation reads:
\begin{equation}
-\lambda^2\nabla^2 H +H +\tau\,\frac{\partial  H }{\partial
t}= \phi_0 \delta({\bm r}-{\bm v}t)\,,\label{TDL1}
\end{equation}
where $H(\bm r,t)$ is the $z$ component of the magnetic field, $\bm r=(x,y)$, and
\begin{equation}
\tau=\frac{4\pi\sigma\lambda^2}{c^2} 
\label{tau}
\end{equation}
is the "current relaxation time", the term used in literature on  TDGL.\cite{Schmid,Kopnin}

     Clearly, the field
distribution described by Eq.\,(\ref{TDL1}) differs from the   solution
which would have existed in the absence of the term $\tau\partial_t  H$,
\begin{equation}
H_0({\bm r},t)=\frac{\phi_0}{2\pi\lambda^2}\,K_0\left(\frac{|{\bm r}-{\bm
v}t|}{\lambda}\right)\,,\label{h0(r,t)}
\end{equation}
which corresponds to translation of the static field distribution with velocity $\bm v$;   $K_0$ is the Modified Bessel Function.

As is done for the diffusion equation,\cite{LL} Eq.\,(\ref{TDL1}) can
be solved by first finding the time dependence of the Fourier transform
$H_{\bm k}$, $\bm k=(k_x,k_y)$:
\begin{equation}
\tau\,\partial_tH_{\bm k} + (1+\lambda^2k^2)H_{\bm k}= \phi_0 \,e^{-i{\bm k}{\bm v} t}\,.
\end{equation}
The general solution of this equation is
\begin{equation}
     H_{\bm k} =  \frac{\phi_0 \,e^{-i{\bm k} {\bm v} t}}{1+\lambda^2k^2-i{\bm k}
{\bm v}\tau} + C e^{-t(1+\lambda^2k^2)/\tau } \,.
\end{equation}
For a stationary solution, the arbitrary constant $C$ is zero.

To find the field distribution in   real space for a constant $\bm v$ it suffices to consider $t=0$:
\begin{equation}
     H({\bm r},t=0) =   \frac{\phi_0 }
{4\pi^2 }\,\int  \frac{d{\bm k}\,e^{ i{\bm k} {\bm r} } }{1+\lambda^2k^2-i
k_x v \tau}\,,
\end{equation}
where $x$ is chosen along the velocity.
 
Integration over $k_x$ is straightforward since the poles of the integrand are
easily found:
\begin{eqnarray}
&&  H\,\frac{2\pi\lambda^2 }{\phi_0} =    e^{-xs/\lambda^2 }
\int_0^{\infty}  \frac{e^{ i k_yy-|x|\eta/\lambda} }{\eta } \lambda \,dk_y\,
  ,\nonumber\\
&& \eta= \sqrt{  1+ \lambda^2k_y^2 +s^2/\lambda^2  }\,,\qquad s= v\tau /2\,.
\label{H-int}
\end{eqnarray}
  Although difficult in general, analytic 
integration over $k_y$  can be done for $x=0$ or $y=0$:
\begin{eqnarray}
   H(0,y) &=&  \frac{\phi_0 }
{2\pi\lambda^2 } \,K_0\left(\frac{|y| }{ \lambda^*}\right)\,,\label{13}\\
  H(x,0) &=&   \frac{\phi_0 }{2\pi\lambda^2 }  \,\exp\left(-\frac{
 x s}{\lambda^2}\right)K_0\left(\frac{|x| }{ \lambda^*}\right), \label{14}\\
  \lambda^*&=&\frac{\lambda}{\sqrt{1+s^2/\lambda^2}}\,.\label{15}
\end{eqnarray}
 If $v=0$, 
  this reduces to the standard static London solution.
\begin{figure}[h]
\includegraphics[width=8cm]{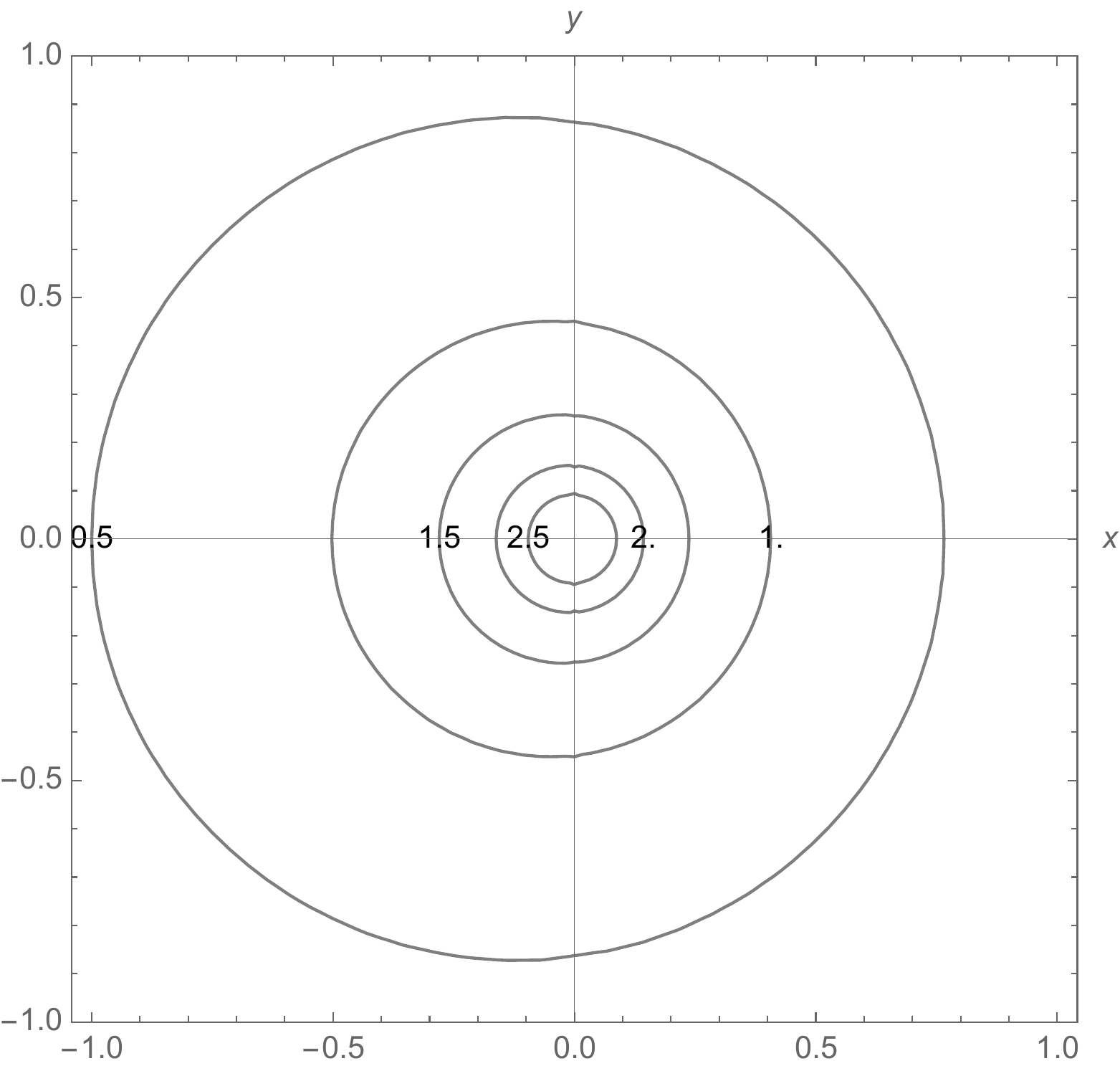}
\includegraphics[width=8cm]{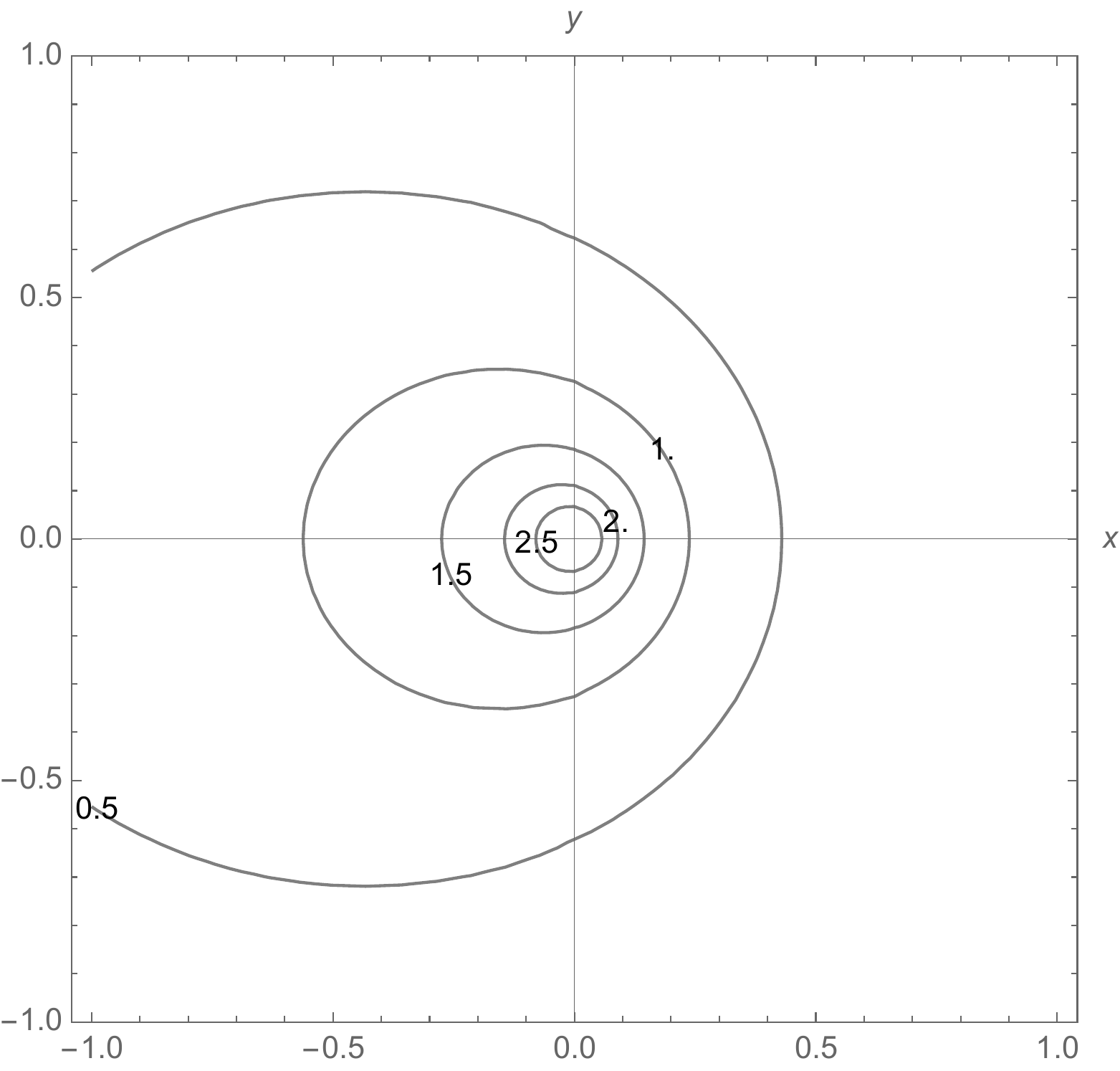}
\caption{ (Color online)  The upper panel: contours of $h(x,y)=$\,const for the parameter $s/\lambda=v/v_c=0.4$. The lower panel: $v/v_c=2$. $x$ and $y$ are in units of $\lambda$.}
\label{fig1}
\end{figure}

Hence,  the field distribution of a moving vortex (a) is not symmetric with respect to $x\to -x$,  (b) is contracted
stronger along the velocity ($x$) than across it ($y$)   (not only the
argument  of the Bessel function $K_0$ is scaled by $\lambda^*<\lambda$, for
 $x>0$  there is an extra factor $e^{-xv\tau/\lambda^2}$ breaking the symmetry $x\to -x$,  and (c)  
  the intervortex interaction in moving system of vortices  differs from that in the static case. 

Physically, the distortion of the field distribution is due to contribution
of the out-of-core normal excitations to  currents of the moving vortex. At small
velocities, the distortion can be disregarded. Indeed, the ratio
\begin{equation}
\frac{s }{\lambda }= \frac{2\pi\sigma\lambda v}{c^2}=\frac{v}{v_c}\,,\quad v_c=\frac{c^2}{2\pi\sigma\lambda}=\frac{2\lambda   }{\tau }\,,
\label{vc}
\end{equation}
where $v_c$ is a crossover value for ``low" and ``high" velocities. At low temperatures $T$, the quasiparticles are nearly absent (for the s-wave symmetry) and $\sigma\approx 0$, whereas $\lambda$ is finite, therefore, the  ratio $v/v_c$ along with the vortex field distortion are small. Hence, the distortion can possibly have an effect at high $T$'s where the conductivity is close to that of the normal phase. Gapless superconductors are exception from this rule, since the normal excitations density of states is close to normal even at low $T$s. 

 The integral (\ref{H-int}) can be evaluated numerically. The upper panel of Fig.\,\ref{fig1} shows contours $h(x,y)=2\pi\lambda^2H/\phi_0=\,$ const for   $v/v_c=0.4$. The lower panel gives $h(x,y)$ for  $ v/v_c=2$; although unrealistic, this example is given to show clearly the distortion of the field distribution of a moving vortex.

\subsection{Extra dissipation by moving vortex}

It should be stressed that the dissipation considered here is due to moving non-uniform distribution of the vortex magnetic field out of the vortex core. This dissipation is usually small relative to  Bardeen-Stephen core dissipation.\cite{Bardeen-Stephen}
The dissipation of interest here is $\sigma E^2$ and 
the electric field $\bm E$ is obtained  with the help of 
Maxwell equations $i({\bm k}\times {\bm E}_{\bm k})_z=- \partial_t\,  H _{z{\bm k}}/c$
and ${\bm k}\cdot{\bm E}_{\bm k}=0$: \cite{remark*}
\begin{eqnarray}
     E_{x{\bm k}} &=&  -\frac{\phi_0v k_xk_y\,e^{-i{\bm k} {\bm v} t}}{ck^2(
1+\lambda^2k^2-i{\bm k} {\bm v}\tau)}  \,,\label{Ex}\\
     E_{y{\bm k}} &=&   \frac{\phi_0v k_x^2\,e^{-i{\bm k} {\bm v} t}}{ck^2(
1+\lambda^2k^2-i{\bm k} {\bm v}\tau)}  \,.\label{Ey}
\end{eqnarray}
It is worth noting that $H_{z{\bm k}}=\phi_0$ at ${\bm k}=0$ (the
flux quantization), i.e., $H_{z,{\bm k}=0}$ does not change in time; therefore
$\bm E_{\bm k=0}$ should be set zero.

The dissipation power per   unit length of the vortex is:
\begin{eqnarray}
w &=&\sigma\int d^2{\bm r}E^2 =\sigma\int \frac{d{\bm
k}}{4\pi^2}\left(|E_{x{\bm k}}|^2+|E_{y{\bm k}}|^2\right)\nonumber\\
&=&\frac{\phi_0^2\sigma v^2}{4\pi^2c^2} \int \frac{d{\bm k}\,k_x^2}
{k^2| 1+\lambda^2k^2-i{\bm k} {\bm v}\tau |^2} \,.
\end{eqnarray}
Note that the contribution to the dissipation due to changing in time order parameter modulus\cite{Schmid,Bardeen-Stephen} is left out in the London approximation.
After integration over ${\bm k}$ directions, the integral  takes the form:
\begin{equation}
\int_0^{\infty}\frac{dk}{k}\left[1-\frac{1+\lambda^2k^2}
{\sqrt{(1+\lambda^2k^2)^2+4k^2s^2}}\right]=
\frac{\pi\lambda^2}{2s^2}\ln\left(1+\frac{s^2}{\lambda^2}\right).
\end{equation}
Thus, we obtain
\begin{equation}
w =\frac{\phi_0^2\sigma v_c^2 }{8\pi c^2\lambda^2}
     \ln\left(1+\frac{v^2 }{v_c^2}\right).
\end{equation}
For $v^2 \ll v_c^2$, we have
\begin{equation}
w =\frac{\phi_0^2\sigma v^2 }{8\pi c^2\lambda^2}\,, \label{W_L}
\end{equation}
so that the quantity
\begin{equation}
\eta_L=\frac{\phi_0^2\sigma   }{8\pi c^2\lambda^2}
\end{equation}
is the London contribution to the drug coefficient (for unit length of a single vortex).
 Bardeen-Stephen
dissipation related to vortex cores corresponds to drag coefficient $\eta_{BS}= \phi_0^2\sigma_n /2\pi c^2\xi^2 $, where $\sigma_n$ is the normal state conductivity. Hence,
 $\eta_L/\eta_{BS}\sim\sigma/\sigma_n\kappa^2$.  Again, this ratio is not necessarily small in gapless materials, see also the remark \onlinecite{remark}.

The opposite limit, $v^2/v_c^2\gg 1$, is hardly realistic at low $T$'s. However, at high $T$'s     both $\lambda$ and the normal excitations conductivity increase when approaching $T_c$, while $v_c$ drops, Eq.\,(\ref{vc}). Hence, the ratio $v^2/v_c^2 $ may become large. One sees that in this case 
 the dissipation is not proportional to $v^2$:
\begin{equation}
w \approx  \frac{\phi_0^2\sigma v_c^2 }{4\pi c^2\lambda^2 }      \ln \frac{v  }{v_c} .
     \label{large v}
\end{equation}
 Therefore, at high temperatures the London dissipation is not analogous to that of the  viscous flow, $w \propto \ln(v/v_c)$.
 
It is worth noting   that there are situations when the vortex velocities are very high. An example is   flux avalanches observed in thin YBCO films with   velocities up to $5\times 10^{6}\,$cm/s.\cite{Leiderer} Also, a very high vortex velocities were recently recorded in Pb films.\cite{Eli}

\section{Dissipation by  moving lattice}

It has been mentioned above that the interaction of moving vortices differs from the static situation. Finding the VL structure of moving VLs is in general complicated, one of the reasons being that the energy is no longer  a thermodynamic potential with a minimum at the structure of the moving lattice even at a constant velocity.  Instead one has to consider the dissipation and use the principle of minimum  entropy production.\cite{Barukh} 
 
The dissipation power per unit volume is given by
\begin{eqnarray}
W= \sigma\frac{B}{\phi_0}\int_{cell} d{\bm r}|E |^2 = = \sigma  \sum_{\bm G\ne 0} |E( {\bm G}) |^2 \,,
\end{eqnarray}
 where the sum is over the reciprocal lattice $\bm G$. The
Fourier components of $\bm E$ are given by Eqs.\,(\ref{Ex}),
(\ref{Ey}) in which $\phi_0$ should be replaced with $B$ and $\bm k\to \bm G$ (see, e.g., Ref.\,\onlinecite{CDK}):
\begin{eqnarray}
W= \frac{\sigma B^2v^2}{c^2}\sum_{\bm G\ne 0}
\frac{G_x^2}{G^2[(1+\lambda^2G^2)^2+G_x^2v^2\tau^2]}\,.
\label{W-lattice}
\end{eqnarray}

In intermediate fields $H_{c1}\ll H\ll H_{c2}$ one disregards 1 relative to a large $\lambda^2G^2$. Besides, even for $v\sim v_c$, the last term in the denominator is small.  
Indeed,  $v_c\tau /\lambda^2 G\sim 1/\lambda G<<1$. 
  However, the   evaluation of $W$ is still a problem because   the VL is affected by  motion. Therefore, one has to figure out first what structure  the moving lattice adopts and then evaluate the dissipation $W $. 

Since the energy can no longer be used as thermodynamic potential with minimum at the moving VL structure, the only way out is to use the principle of ``minimum entropy production", i.e. to minimize the dissipation power $W$. The problem of $W $ minimization  with respect to different VL structures is challenging: the VL cell is determined by two lattice vectors under the restriction of the flux quantization, in other words, one has to minimize the sum (\ref{W-lattice}) with respect to   variation of three parameters. Unfortunately, this sum has many local minima, results may depend on initial guesses, and the numerical problem becomes massive.

One can pose a less ambitious question. 
The calculations based on  TDGL  near $H_{c2}(T)$ for clean materials have shown that when the VL moves fast enough, it   has one of the unit cell   vectors along the velocity $\bm v$. \cite{Barukh} This result   holds in the presence of disorder as well, where VL adopts this structure at large velocities of the flux flow.\cite{Kosh-Vin} The TDL has an advantage of applicability at all temperatures and in fields well under $H_{c2}$. Hence, it is of interest to see whether a relatively simple TDL gives  results compatible with previous work. 
\begin{figure}[h]
\includegraphics[width=8cm]{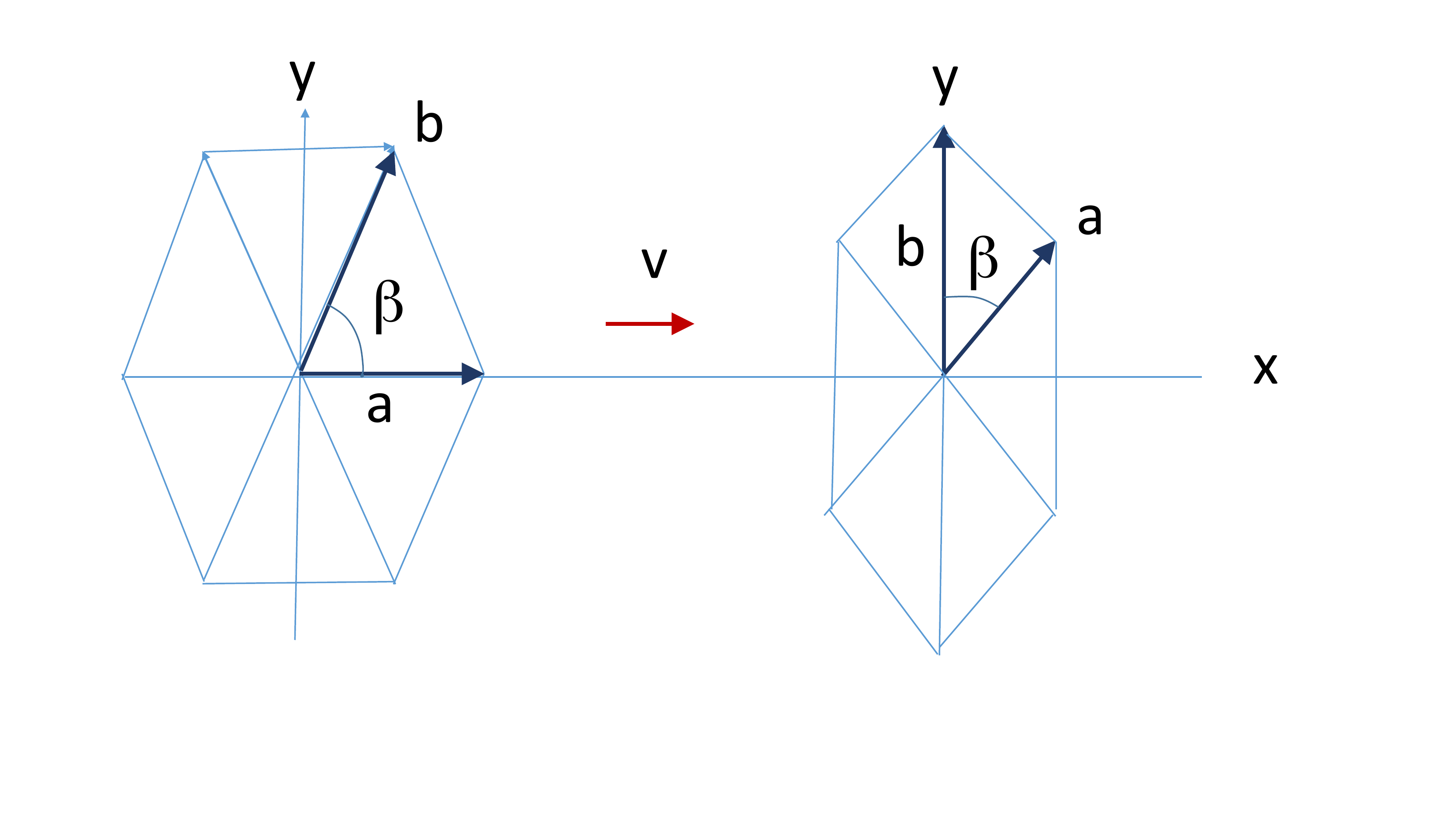}
\caption{ (Color online) Structures $A$ at the left and $A^\prime$ at the right.}
\label{fig2}
\end{figure}

Below,  two structures of moving VLs are compared: the structure $A$ with one of the cell vectors parallel to $\bm v$ and $A^\prime$ where one of the cell vectors is normal to $\bm v$. Both structures are assumed to consist of isosceles triangles, as shown in Fig\,\ref{fig2}; this is imposed for $A^\prime$ by necessity to have the $x$ axis as a symmetry plane; for $A $ the $y$ axis is assumed to be a  symmetry plane. One readily obtains the reciprocal lattice for $A$:
\begin{eqnarray}
 G_x=\frac{2\pi}{\lambda} \sqrt{\frac{ h\tan \beta}{2}}\,m,\quad
  G_y=\frac{2\pi}{\lambda} \sqrt{\frac{2 h}{\tan \beta}}\left(n-\frac{m}{2}\right),\qquad 
\label{GxyA}
 \end{eqnarray}
where $m,n$ are integers, $h=B\lambda^2/\phi_0$ and  $\beta$ is the angle between unit cell vectors, see Fig\,\ref{fig2}. The reciprocal lattice for the structure $A^\prime$ is:
\begin{eqnarray}
  G_x^\prime=\frac{2\pi}{\lambda}\sqrt{\frac{2 h}{\tan \beta}} \left(m-\frac{n}{2}\right) ,\quad
  G_y^\prime=\frac{2\pi}{\lambda}\sqrt{\frac{ h\tan \beta}{2}}n.\qquad 
\label{GxyB}
 \end{eqnarray}
These structures are determined by two parameters, $h$ and $\beta$ (the cell area and the angle between cell vectors). 
 
\begin{figure}[h]
\includegraphics[width=7cm]{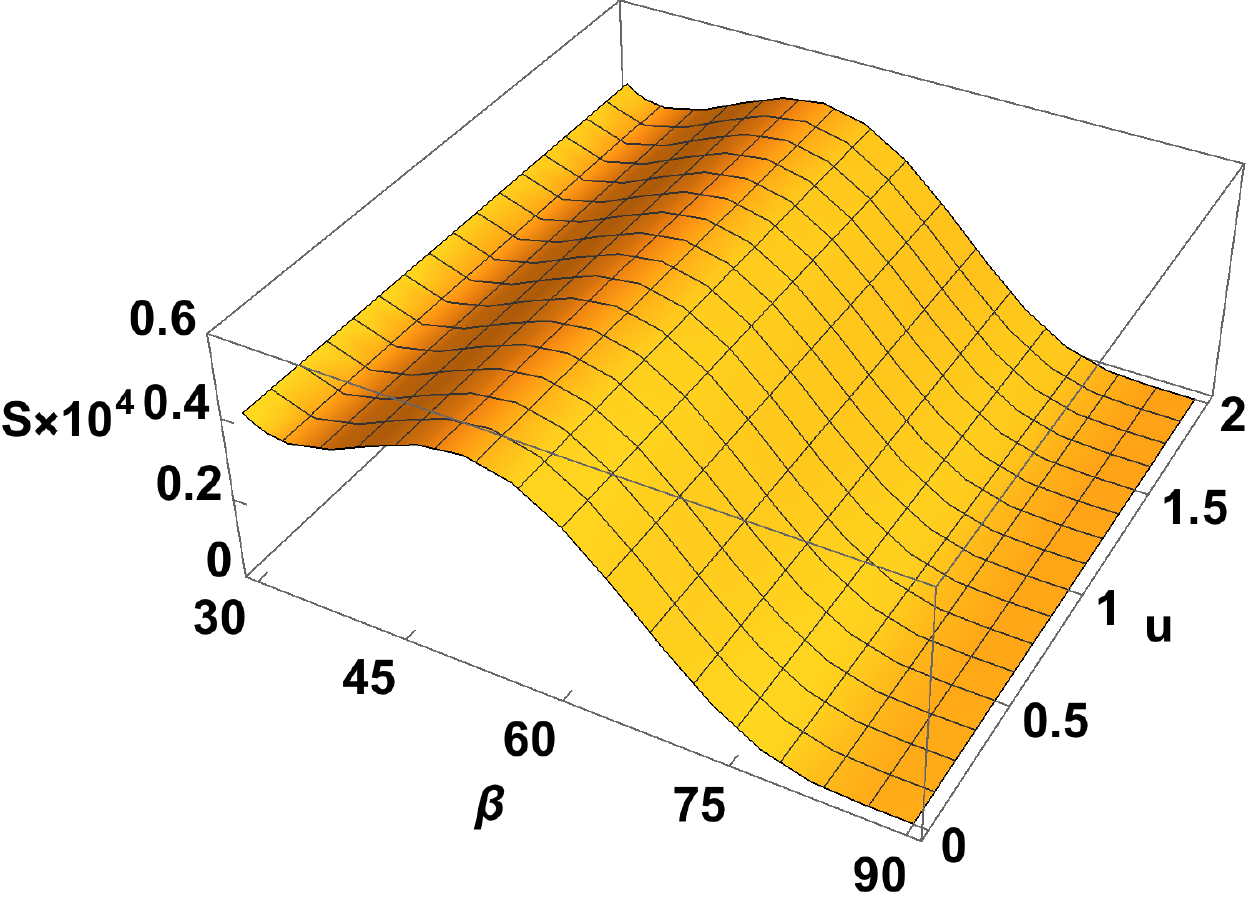}
\caption{ (Color online) The sum  $S\times 10^4$ vs angle $\beta$ in degrees and  velocity $u=v/v_c$ for $h=3$ and $\kappa=10$.  $S$ is nearly $u$ independent in the interval of velocities chosen.}
\label{fig3}
\end{figure}

To compare the   dissipation  for these two VLs, one has to evaluate the sum (\ref{W-lattice}). This sum is slowly convergent, so that, when calculated numerically,  it will depend on a  summation domain chosen for $m$ and $n$. On the other hand, within the London approach, there is no sense to extend summation to  $G>1/\xi$. To make this truncation smooth, one adds to the summand a factor $e^{-G^2\xi^2}$ and calculates the dimensionless sum
\begin{eqnarray}
S = \sum_{\bm g\ne 0}
\frac{g_x^2 \,e^{-g^2/\kappa^2}}{g^2[(1+g^2)^2+4 g_x^2u^2]}\,,
\label{S}
\end{eqnarray}
where $\bm g=\lambda\bm G$, $u=v/v_c$, and $\kappa=\lambda/\xi$ is the GL parameter. It turns out that the quantity $S(\beta,h,u)$  is nearly velocity independent    in the range $0<u<2$.  
Numerically evaluated $S(\beta,u)$ for a fixed $h$ is shown in Fig.\,\ref{fig3}.
Hence, the London contribution to the dissipation power, 
\begin{eqnarray}
W=\frac{\sigma B^2}{c^2}\,S\,v^2 \,,
\label{W-S}
\end{eqnarray}
is  proportional to $v^2$ as in a viscous flow. The drag coefficient, however, depends on $S$, i.e. on the VL structure (the angle $\beta$  and the field $h$). 
\begin{figure}[t]
\includegraphics[width=7cm]{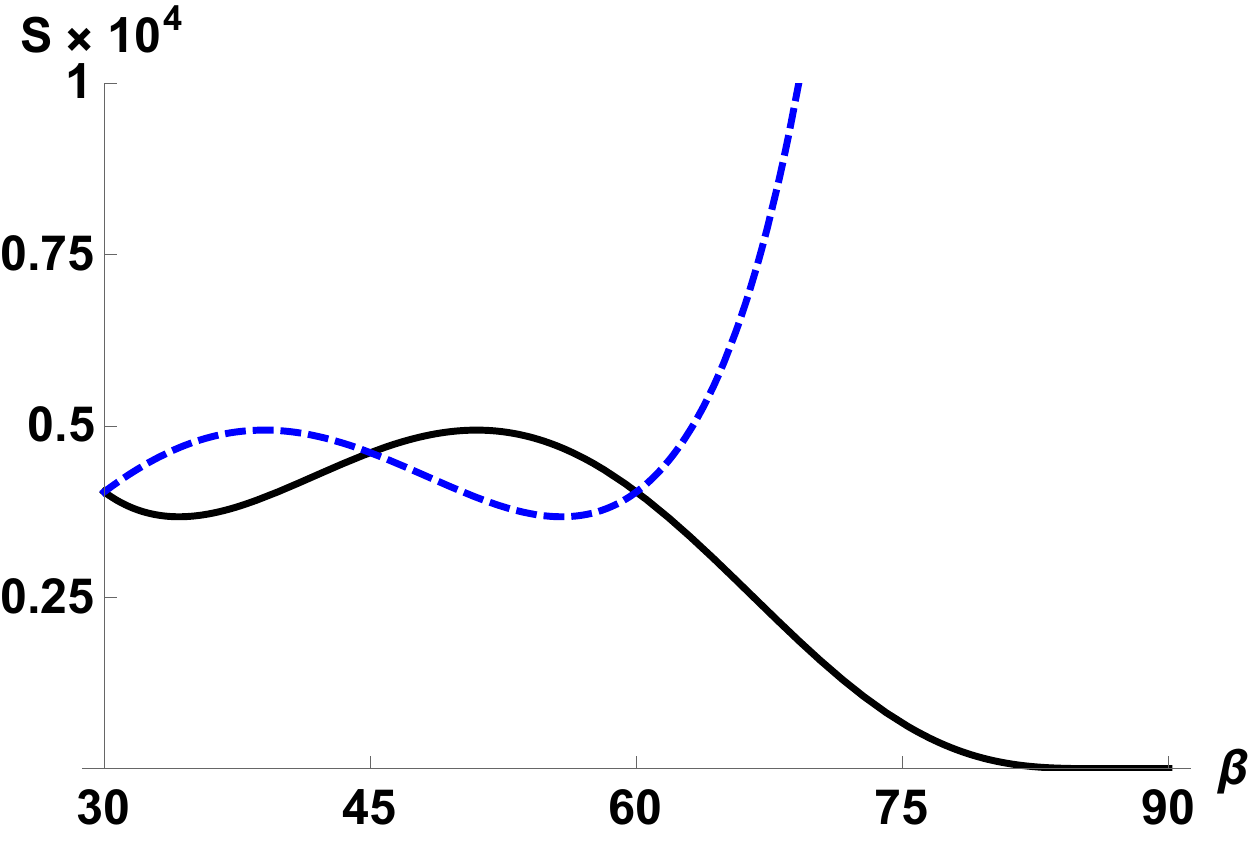}
\caption{ (Color online)  The sum  $S$ vs angle $\beta$   for $h=3$ and $\kappa=10$. The solid curve is for the structure $A$, the dashed one is for $A^\prime$. Clearly, for $\beta >60^\circ$ the dissipation of the structure $A$ is less than that of $A^\prime$.}
\label{fig4}
\end{figure}

The quantity $S$   is plotted in Fig.\,\ref{fig4} for $h=3$, $\kappa=10$, $u=0.5 $ for structure $A$ (solid line) and $A^\prime$ (dashed line).
For $\beta > 60^\circ $, the dissipation is clearly the lowest when one of the unit cell vectors is parallel to the velocity, the structure $A$, the result   obtained in Ref.\,\onlinecite{Barukh} with time-dependent GL theory. This demonstrates that the   time-dependent London model works qualitatively well, shortcomings (the cores are out) of the London approach  notwithstanding and with added bonus of arbitrary temperatures and simplicity. 
\begin{figure}[h]
\includegraphics[width=7.cm]{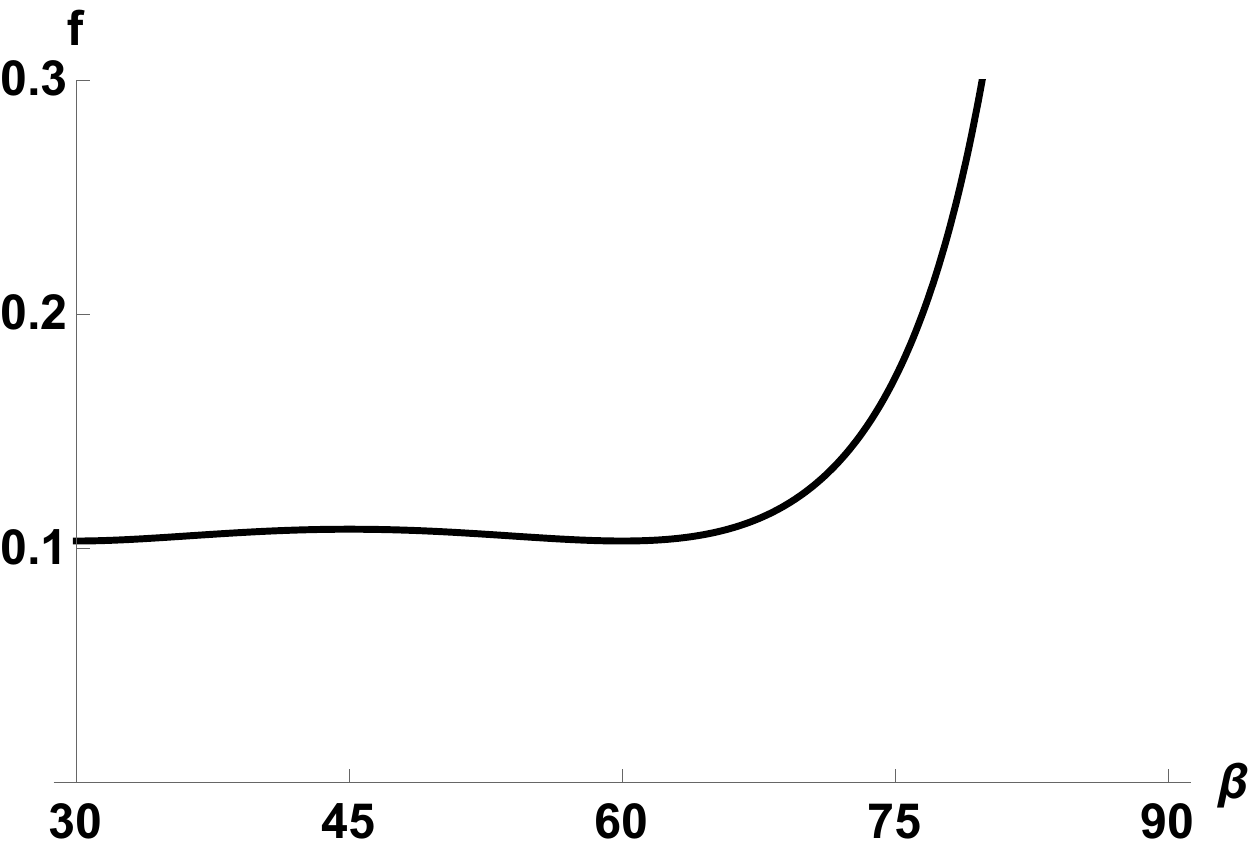}
\caption{ (Color online)  Dimensionless energy $f =  (8\pi\lambda^4/\phi_0^2)\,F$ of the structure $A$   vs angle $\beta$ for $h=3$, $\kappa=10$.}
\label{fig5}
\end{figure}

It is worth noting  that, as shown in Fig.\,\ref{fig4}, the London drag vanishes when $\beta\to 90^\circ$ for the structure $A$.  This means that the principle of minimum entropy production pushes the system to a structure made of isosceles triangles with a shrinking base $a$ and large distance $\approx b$ between rows parallel to $\bm v$, see the left panel of Fig.\,\ref{fig2}. Such a structure would look as  system  of vortex chains parallel to $\bm v$ with inter-chain separation $b\sin\beta \gg a$. Physically, dissipation for this structure is low   because   in dense chains the $x$ dependence of the magnetic field distribution is  weak. When such a structure moves along $x$, the time derivative of the field in the laboratory frame $\partial_t h=-v\,\partial_x h\to 0$, i.e. the induced electric field would also be  small. One can say that the system has a tendency to move along the channels (chains)  with low dissipation. 

The trend to transform VL to  a chain structure, i.e. to larger $\beta$, will   however  be opposed by increase of the VL energy for   strong deviations from the static hexagonal arrangement. The energy of interacting vortices within the structure $A$ can be estimated for intermediate fields:  \cite{deGennes} 
\begin{equation}
F =  \frac{B^2}{8\pi}\sum_{\bm G\ne 0}\frac{1}{1+\lambda^2G^2}\approx \frac{\phi_0^2}{8\pi\lambda^4}h^2\sum_{\bm g\ne 0}\frac{e^{-g^2/\kappa^2}}{g^2}\,. 
\label{F}
\end{equation}

The quantity $f =  (8\pi\lambda^4/\phi_0^2)\,F$   evaluated numerically is shown in Fig.\,\ref{fig5} for $h=3$, $\kappa=10$. 
A sharp increase of $f(\beta)$ for $\beta > 60^\circ$ and divergence at $\beta\to\pi/2$ suggest  that $\beta$ cannot approach $\pi/2$ since the VL energy there will exceed the condensation energy.  For a clean material at low temperatures and intermediate fields, one can roughly estimate the energy of a hexagonal VL as
\begin{equation}
F_0 \approx \frac{\phi_0B }{32 \pi^2\lambda^2}\ln\frac{H_{c2}}{B}\,, \quad \lambda^{-2}=\frac{8\pi e^2 N(0)v^2_F}{3c^2}\,, 
\label{F0}
\end{equation}
where $e$ is the electron charge, $v_F$ is the Fermi velocity, and $N(0)$ is the density of states per spin. The condensation energy in this case is $F_c=N(0)\Delta^2(0)/2$. One easily obtains that the ratio $F_0/F_c\sim B/H_{c2}$ is not a very small number. That means that formally the energy of a deformed VL can easily approach the condensation energy.

As an example, equating the ratio of the condensation energy $H_c^2/8\pi=\phi_0^2/64\pi^3\lambda^2\xi^2$ at high temperatures to the energy (\ref{F}) one obtains $f(\beta)=\kappa^2/8\pi^2\approx 1.27$ for parameters of Fig.\,\ref{fig5}. One redily finds numerically that this corresponds to the maximum possible $\beta\approx 86^\circ$, which corresponds to the  separation between vortices within the chain $a\approx 0.216\, \lambda$ whereas the interchain distance $b\approx 1.54\,\lambda$. 

\section{Summary}

Thus, it is demonstrated that the formalism of time-dependent London equations can be employed to consider dynamic problems of type-II superconductivity provided the order parameter modulus can be considered  constant. The linear TDL approach is much simpler than, e.g., the non-linear time-dependent GL. As in the static case, TDL provides a simple all-temperatures tool to address such problems as moving vortices and vortex lattices. 

The TDL   is based on the notion that in $t$ dependent situations, the current consists of normal and superconducting parts, Eq.\,(\ref{current1}). What follows is a diffusion type equation (\ref{TDL}) for the magnetic field. It is shown  that the field distribution of a moving vortex is distorted, it is contracted and looses cylindrical symmetry, Eqs.\,(\ref{13})--(\ref{15}) and Fig.\,\ref{fig1}.

The $t$ dependent field distribution of a moving vortex gives rise to an extra dissipation not included in the Bardeen-Stephen core dissipation. The London dissipation is usually  small   but it may constitute a considerable part of the core dissipation in gapless materials. It is also worth noting that as $T\to T_c$, the velocity separating slow and fast motion $v_c=c^2/2\pi\sigma\lambda$ becomes small. Then if $v>v_c$, the London dissipation is $\propto\ln(v/v_c)$, i.e. the analogy with slow viscous flow  is lost.

Distortions of the field distribution of single moving vortices lead to a distorted inter-vortex interactions and therefore to a change in the vortex lattice structure. 
Effects of disorder were considered in Ref.\,\onlinecite{Kosh-Vin}; it turned out that at large velocities the moving VL adopts the structure with one of the lattice vectors along the velocity, the same result as in clean systems studied within   TDGL in Ref.\,\onlinecite{Barukh}.  This effect was seen in a few experiments. \cite{Yaron,March,Pardo,Forgan} Employing the principle of minimal dissipation in a stationary state, it is shown  that TDL can reproduce this result without the temperature restrictions of TDGL.  \\

The author is grateful to L. Bulaevskii, E. Zeldov, R. Mints,  A. Gurevich, and H. Suderow for useful discussions. The work was supported by the U.S. Department of Energy, Office
of Science, Basic Energy Sciences, Materials Sciences and
Engineering Division. The Ames Laboratory is operated for
the U.S. DOE by Iowa State University under Contract No.
DE-AC02-07CH11358.


\references


\bibitem{London}F. London {\it Superfluids}, vol.1, Dover, New York, 1950.

\bibitem{Clem-Coffey}M.W. Coffey and J.R. Clem, \prb. {\bf 46}, 11757 (1992).


\bibitem{Yaron}U. Yaron et al., Phys. Rev. Lett. {\bf 73}, 2748 (1994).

\bibitem{March}M. Marchevsky, J. Aarts, P. Kes, and M. Indenbom, Phys. Rev.
Lett. {\bf 78}, 531 (1997).

\bibitem{Pardo}F. Pardo, F. de la Cruz, P.L. Gammel, E. Bucher, and D.J.
Bishop,   Nature (London) {\bf 396}, 348 (1998).

\bibitem{Forgan}D. Charalambous, P. G. Kealey, E. M. Forgan, T. M. Riseman, M.
W. Long, C. Goupil, R. Khasanov, D. Fort, P. J. C. King, S. L.
Lee, and F. Ogrin, Phys. Rev. B, {\bf 66}, 054506 (2002).

 \bibitem{Kosh-Vin}A. Koshelev, V. Vinokur, \prl {\bf 73}, 3580 (1994).
 
 \bibitem{Barukh}  D. Li,  A. M. Malkin,  and B. Rosenstein, \prb {\bf 70}, 214529 (2004).

 \bibitem{Tinkham} M. Tinkham, {\it Introduction to Superconductivity}, McGraw-Hill, New York, 1996.

 \bibitem{Schmid}A. Schmid, Phys. Kondens. Materia {\bf 5}, 302 (1966).

\bibitem{Kopnin} N. B. Kopnin, {\it{Theory of Nonequilibrium
Superconductivity}} (Oxford University Press, New York, 2001).

     \bibitem{LL}L.D. Landau and E.M. Lifshitz, {\it Fluid Mechanics}, Pergamon Press, 1987.
 

\bibitem{Bardeen-Stephen}J. Bardeen  and M. J. Stephen. Phys. Rev. {\bf 140}, A1197 (1965).  

\bibitem{remark*}Within TDGL, to account for possible transformation of the normal current within the core to the supercurrent outside, only the total current  satisfies   div$(\bm J_n +\bm J_s)=0$, where $\bm J_n $ is due to normal excitations and $\bm J_s$ is the supercurrent. Within TDL, since we are away of cores, the conditions  div$ \bm J_n =0$ and div$ \bm J_s =0$ should be satisfied $\it separately$.

\bibitem{remark} In a series of publications, Maeda's group reports  the core conductivity smaller than $\sigma_n$, see e.g., 
T. Okada, H. Takahashi,  Y. Imai,  K. Kitagawa,  K. Matsubayashi,  Y. Uwatoko,  and A. Maeda, \prb {\bf 86}, 064516 (2012).  This may push up the ratio $\eta_L/\eta_{BS}$. 

\bibitem{Leiderer}B. Biehler, B.-U. Runge, P. Leiderer, and R. G. Mints,
Phys. Rev. B {\bf 72}, 024532 (2005). 

\bibitem{Eli}L. Embon, Y. Anahory, \v{Z}.L. Jeli\'{c}, E.O. Lachman, Y. Myasoedov, M.E. Huber, G.P. Mikitik,
A.V. Silhanek, M.V. Milosevi\'{c}, A. Gurevich, and  E. Zeldov, Nature Communications, {\bf 8}, 85 (2017). 

\bibitem{CDK}L. J. Campbell, M. M. Doria, and V. G. Kogan, \prb {\bf 38}, 2439
(1988).

\bibitem{deGennes}P. deGennes,  {\it Superconductivity of metals and alloys}, Benjamin, New York, 1966.



 \bibitem{remark3}It is of interest to mention that   VL reorientation to have one of the lattice vectors along the velocity has been observed not only in flux-flow experiments, but surprisingly in the flux creep data as well: E. Herrera, J. Benito-Llorens, U. S. Kaluarachchi, S. L. Bud'ko, P. C. Canfield, I. Guillamón, and H. Suderow, \prb {\bf 95}, 134505 (2017). Average creep velocities are exceedingly small, so that the model of this paper $\it {per\,\, se}$ cannot be applied.
\end{document}